\documentclass[12pt,a4paper]{article}

\usepackage{pstricks}
\usepackage{graphicx}
\usepackage[latin2]{inputenc}
\usepackage{epic}
\usepackage{latexsym}
\usepackage{epsf}
\usepackage{epsfig}

\newcommand{\eref}[1]{eq.\ (\ref{e.#1})} 
\newcommand{\erefn}[1]{ (\ref{e.#1})}
 
\newcommand{\aref}[1]{\ref{a.#1}}
\newcommand{\sref}[1]{Section \ref{s.#1}}
\newcommand{\cref}[1]{Chapter \ref{c.#1}}

\def\nn{\nonumber \\}  
\newcommand{\nl}{& \nonumber \\ &}
\newcommand{\bnl}{\right . & \nonumber \\ & \left .}
\def\ll{\label}
\def\ds{\displaystyle}

\def\kahler{K\"ahler\hspace{0.1cm}}
\def\sugra{supergravity\hspace{0.1cm}}

\def\beq{\begin{equation}} 
\def\eeq{\end{equation}} 
\def\bea{\begin{eqnarray}}  
\def\eea{\end{eqnarray}}  
\def\ba{\begin{array}}  
\def\ea{\end{array}}   
\def\bi{\begin{itemize}}  
\def\ei{\end{itemize}}  
\def\be{\begin{enumerate}}  
\def\ee{\end{enumerate}}  
\def\beq{\begin{equation}}  
\def\eeq{\end{equation}}  
\def\bc{\begin{center}}
\def\ec{\end{center}}

\def\ca{{\mathcal A}}
   
\def\cl{{\mathcal L}}  
\def\cf{{\mathcal F}}   
  
\def\co{{\mathcal O}}

\def\kr{k \pi R}

\def\dO{\Delta \Omega}
\def\intdk{\int {d^4 k \over (2 \pi)^4}}

\def\pa{\partial}

\def\rt{\sqrt{2}}

\def\hc{{\rm h.c.}}

\def\ov{\overline}

  \def\eps{\epsilon}

\begin{document}

\pagestyle{empty}
\begin{flushright}
hep-th/0502072\\
DESY-05-024
\end{flushright}
\vskip 1.5cm

\begin{center}
{\huge On the one-loop \kahler potential  in five-dimensional brane-world supergravity}
\end{center}
\vspace*{5mm} \noindent
\vskip 0.5cm
\centerline{\bf Adam Falkowski}
\vskip 1cm
\centerline{\em Institute of Theoretical Physics, Warsaw University}
\centerline{\em ul.\ Ho\.za 69, PL-00-681 Warsaw, Poland}
\vskip 0.5cm
\centerline{\em Deutsches Elektronen-Synchrotron DESY}
\centerline{\em Notkestrasse 85, 22607 Hamburg, Germany}
\vskip .5cm
\centerline{\tt \small E-mail: afalkows@mail.desy.de}
\vskip1cm
\centerline{\bf Abstract}
\vskip .3cm
We present an on-shell formulation of 5d gauged supergravity coupled to chiral matter multiplets localized at the orbifold fixed points. The brane action is constructed via the Noether method. In such set-up  we compute one-loop corrections to the \kahler potential of the effective 4d supergravity and compare the result with  previous computations based on the off-shell formalism. The results agree at lowest order in brane sources, however at higher order there are differences. We explain this discrepancy by an ambiguity in resolving singularities associated with the presence of infinitely thin branes.  
\vskip .3cm

\newpage
\pagestyle{plain}

\section{Introduction}

Supersymmetry breaking and its mediation to the observable sector is one of the most important problems in physics beyond the Standard Model. An acceptable theory of supersymmetry breaking is strongly constrained by the observed features of the low-energy physics. Spontaneous breaking must occur in a hidden sector and must be transmitted to the observable sector via non-renormalizable operators. Gravity mediation is an attractive and economical possibility, but it is well known that generic models face the flavor problem.  

It has been noted in ref. \cite{rasu0} that  a spatial separation of the hidden and observable sectors  brings new elements into the mechanism of gravity mediation. The simplest set-up that provides for such sequestering is that of five-dimensional (5d) supergravity compactified on an orbifold in which the chiral matter of the observable and hidden sectors is localized on the two  different boundaries of the fifth dimension.  
In the minimal set-up with no matter fields in the bulk the tree-level K\"ahler potential of the effective 4d theory has a special structure that results in absence of tree-level transmission of supersymmetry breaking. 
Supersymmetry breaking is  then transmitted to the observable sector at one-loop level by (flavor-blind) gravitational interactions. One mechanism that can operate here is  anomaly mediation \cite{rasu0,gilumu}. Besides, there is always a contribution from integrating out the Kaluza-Klein (KK) tower of the supergravity multiplet. Its effect on the low-energy phenomenology can be summarized as a correction to the tree-level \kahler potential of the 4d effective supergravity. 
This one-loop correction was computed in refs. \cite{ghri,bugago,rascst,grrasc}. Unlike in four-dimensions, the contact terms between the hidden and the observable sectors generated by gravity loops are finite and calculable. Therefore 5d supergravity models open a possibility of constructing a realistic and predictive theory of soft terms. 
See also \cite{alle} for other studies of 5d brane-world supergravity. 

 Given the important role of gravitational loop corrections it is advantageous to study them in a somewhat different setting.
The brane-world action considered in refs. \cite{ghri,bugago,rascst,grrasc} was based on an off-shell formulation of 5d supergravity. In this paper we point out that the physics of 5d brane-worlds can be studied in a simpler set-up of on-shell supergravity.  Using the Noether procedure, we construct a locally supersymmetric action for an $N=1$ chiral multiplet confined to a 4d brane and coupled to 5d gauged supergravity in the bulk. In such set-up  we compute the one-loop corrections to the  \kahler potential and compare it with the previous results.

In principle, the Noether method is less powerful than the off-shell formalism of ref. \cite{zu} or the superconformal tensor calculus of ref. \cite{fukuoh}. Still, we will argue that it has several advantages. Firstly, it is obviously less involved. The number of fields is reduced  and no advanced superspace techniques are needed for constructing the action. We will also see that one-loop computations are considerably simplified in this set-up.   
Secondly, it facilitates the treatment of singularities associated with the presence of infinitely thin (delta-like) branes.  In the off-shell formulation integrating out auxiliary fields generates singular terms in the brane action. These singular terms can be avoided in our Noether formulation.
Furthermore, working in our set-up  we will  notice certain ambiguity in defining the brane-world action that is connected to arbitrariness in resolving the singular behaviour of bulk fields near the branes. In certain circumstances, namely when brane sources are large compared to the compactification scale,  this  ambiguity may also affect low-energy observables.
Finally, the procedure can be readily generalized to higher-dimensional spacetimes where an off-shell formulation of supergravity does not exist (for example, similar method was used for coupling 10d brane to  11d supergravity in  the Horava-Witten model \cite{howi}).

The paper is organized as follows. In \sref{bwl} we construct an on-shell action for  an $N=1$ chiral multiplet coupled to 5d supergravity. In \sref{elt} we derive the tree-level \kahler potential describing the dynamics of the low energy degrees of freedom in this set-up. In \sref{olc} we compute the one-loop correction to the \kahler potential and in \sref{akp} we comment on the differences with the  previous works. In this paper we restrict to studying technical issues associated with the Noether construction and one-loop computation. 
 Phenomenological issues, like moduli stabilization or determination of soft breaking terms, are left for future publications.  

\section{Five-dimensional brane-world action}
\label{s.bwl}

In this section we construct a locally supersymmetric action for an $N=1$ chiral multiplet confined to a 4d brane and coupled to  5d supergravity in the bulk. We use the Noether method. That is, starting with a globally supersymmetric action for the brane multiplet we systematically add new terms to the action and supersymmetry transformations until the bulk+brane set-up becomes locally supersymmetric. We first work out all necessary zero- and two-fermion terms such that all 
 two-fermion supersymmetric variations  of the brane action cancel.  The next step is to  determine four-fermion terms
from cancellation of four-fermion variations. In fact, the latter step will not be presented here, but see \cite{aa_master}. It should be stressed however, that at the two-fermion level the brane action we construct is locally supersymmetric to arbitrary power in $1/M_5$ expansion, where $M_5$ is the  5d Planck scale.   

The 5d bulk contains $N=2$ supergravity multiplet\footnote{Our notation and conventions are summarized in \aref{nc}} $(e_\alpha^a, \psi_\alpha, \ca_\alpha)$. 
For the flat (ungauged) 5d supergravity the action up to four-fermion terms reads \cite{chni}
\bea & \ds
\ll{e.5da}
\cl = M_5^3 e_5 \left[ 
{1 \over 2} R_5 
- i \ov{\psi_\alpha} \Gamma^{\alpha\beta\gamma} D_\beta \psi_\gamma
- {1 \over 4} \cf_{\alpha\beta} \cf^{\alpha\beta}
+ {1 \over 6 \sqrt{6}} \epsilon^{\alpha\beta\gamma\delta\epsilon}
 \ca_\alpha  \cf_{\beta\gamma} \cf_{\delta\epsilon}
\bnl
+ {3 i \over 4 \sqrt{6}}
\ov{\psi_\alpha} \Gamma^{\alpha\beta\gamma\delta} \psi_\beta  \cf_{\gamma\delta}
+ {3 i \over 2 \sqrt{6}} \ov{\psi_\alpha}\psi_\beta \cf^{\alpha\beta}
\right ] \, ,
\eea
while the supersymmetry transformations, up to three-fermion terms in $\delta \psi$ are given by
\bea
\ll{e.5ds}
\delta e_\alpha^a &=& {i \over 2}  \ov{\psi_\alpha} \Gamma^a \epsilon + \hc \, , 
\nn
\delta \psi_\alpha &=& D_\alpha \epsilon 
- {1 \over 4 \sqrt{6}}
(\Gamma_{\alpha\beta\gamma} - 4 g_{\alpha \beta} \Gamma_\gamma) \epsilon
 \cf^{\beta\gamma} \, , 
\nn
\delta \ca_\alpha &=& -{i\sqrt{6}\over 4}  \ov{\psi_\alpha} \epsilon + \hc \, . 
\eea
 
 The fifth dimension is the orbifold $S_1/Z_2$ parametrized by 
$x_5 \in [-\pi R, \pi R]$ 
with $Z_2$ acting as 
$x_5 \to -x_5$.
 Under $Z_2$ the field components
 $e_\mu^m$, $e_5^5$, $\ca_5$, $\psi_\mu^+\equiv P_R \psi_\mu$ and $\psi_5^+ \equiv P_L \psi_5$ are even, $\psi(-x_5) = \psi(x_5)$, 
while $e_\mu^5$, $e_5^m$, $\ca_\mu$, $\psi_\mu^-\equiv P_L \psi_\mu$ and $\psi_5^- \equiv P_R \psi_5$ are odd,  $\psi(-x_5) = -\psi(x_5)$. 
At the orbifold fixed point $x_5 = 0$ we couple the $N=1$ chiral multiplet  $[Q_0, \psi_Q]$ ($P_L \psi_Q = \psi_Q$).  
Of course the action for the matter  $[Q_\pi, \psi_{Q_\pi}]$ at the orbifold fixed point $x_5 = \pi R$ can be constructed analogously. 
The starting point for the Noether method is the  action  
\beq
\label{e.gsl}
\cl_{1} = e_4 \delta(x_5) \left \{ \pa_\mu Q_0^\dagger \pa^\mu Q_0 + i \ov \psi_Q \gamma^\mu D_\mu \psi_Q \right \}
\eeq 
invariant under  global supersymmetry  transformations
\beq  
\label{e.gst}
\delta Q_0 = {1 \over \rt} \ov \eps \psi_Q \qquad
\delta \psi_Q  = - {1 \over \rt} i \gamma^\mu \pa_\mu Q_0 \, \eps \, .
\eeq
Under the transformations \erefn{gst}  but with  $\eps$  depending on the 4d coordinates $x_\mu$ the lagrangian of \eref{gsl} transforms as 
$\delta \cl = \pa_\mu \eps \, j^\mu$, 
where $j^\mu$ is the Noether current of global supersymmetry (the supercurrent).  In order to cancel this variation we need to couple  one linear combination $\psi_\mu$ of  the two bulk gravitinos $\psi_\mu^+(0)$, $C \ov {\psi_\mu^-}^T(0)$ to the the supercurrent and identify the parameter $\eps$ with the corresponding combination of the two bulk supersymmetry transformation parameters,  $\eps_+(0)$ and  $C \ov {\eps_-}^T(0)$. In absence of brane sources for the gravitinos  $\psi_\mu^-(0) = 0$ and thus we choose $\psi_\mu = \psi_\mu^+(0)$. However, if such sources are present some other combination of the gravitinos couples to the brane matter. This is for example the case when gravitino brane mass terms are present, see the discussion in \aref{bsp}.

Thus we add to the brane action the so-called Noether term,
\beq
\label{e.nt}
\cl_{2} = - {1 \over \rt} e_4 \delta(x_5)  
\ov \psi_Q \gamma^\mu  \gamma^\rho \pa_\rho Q_0 \psi_\mu + \hc \, .
\eeq  
At the level of two-fermion terms there are still variations to be canceled. One originates from varying the gravitino in \eref{nt},
$ \delta \psi_\mu \sim \cf_{\mu 5}$,  
 the other from variation of the vielbein in the kinetic terms of \eref{gsl}.    It turns out that the necessary modifications of the brane action can be concisely summarized as the redefinition of  the graviphoton field strength. Namely, in the 5d bulk action \erefn{5da} and supersymmetry transformations \erefn{5ds} we replace 
$\cf_{\mu  5}$ with   $\hat \cf_{\mu 5}$ defined as 
\bea \ds  &
\label{e.hfs}
 \hat \cf_{\mu  5} =  \cf_{\mu  5}+  {1 \over M_5^3} \delta(x_5) j_\mu^0 \, ,
\nl \ds 
j_\mu^0 =  {i \over \sqrt 6}\left [ 
Q_0^\dagger \pa_\mu Q_0 -\pa_\mu Q_0^\dagger Q_0 
+ {i \over 2} \ov\psi_Q \gamma_\mu \psi_Q  
\right ]   \, , 
\eea
and modify the transformation law of the  graviphoton by
\beq
\ll{e.gt}
\delta \ca_5 = {i \over \sqrt{12}}\delta(x_5) \ov \psi_Q \eps Q_0 + \hc  \, .
\eeq
In other words we modify the Bianchi identity for the graviphoton field strength such that
$\pa_{[\mu} \hat \cf_{\nu] \, 5} = {2 i \over \sqrt 6 M_5^3} \delta(x_5) \pa_{[\mu} Q_0^\dagger \pa_{\nu]} Q_0$. 
The replacement $\cf \to \hat \cf$ generates  singular $\delta^2$ terms in the brane action. However such singular terms are absent in the low energy effective theory after integrating out the graviphoton $\ca_\mu$. The reason for this is precisely the fact that the singular $\delta^2$ terms match the full square structure inside the graviphoton field strength. In the 5d setup these singular terms provide for necessary  counterterms to cancel divergences in certain one-loop diagrams \cite{mipe}. No other singular terms arise in this construction.\footnote{The Noether construction of brane action in flat 5d supergravity was also pursued in ref. \cite{digeko} but their results differ from ours, notably by the absence of the full square structure.}   

The on-shell action we derived by the Noether method differs from the brane action obtained in the off-shell formalism after eliminating the auxiliary fields \cite{rascst}. In particular the kinetic terms of the gravity multiplet do not couple to the brane here. 
In \sref{akp} we will discuss this issue more carefully and argue that the two formalisms are related by a redefinition of the 5d degrees of freedom. Note that the choice of variables we use here is very convenient, as no singular 
$\delta^n$ terms are present in the on-shell action (except for the $\delta^2$ fitting the full square).   

At this point all the two-fermion variations are canceled. More involved calculations are needed to work out four-fermion terms in the brane action as well as three-fermion modifications of the supersymmetry transformation laws. They are not necessary for the following analysis and will not be presented here.\footnote{The complete action is given in  ref. \cite{aa_master}. See also this reference for coupling of $N=1$ gauge multiplets on the brane.} Indeed, we shall see that the form of the low energy 4d supergravity (including one-loop corrections) can be read out from the terms we have already derived. 
 Note also that once tree-level effective supergravity is known all the three- and four-fermion terms can be easily inferred by matching with the canonical 4d supergravity lagrangian.

This construction of the brane world action can be carried over to the case of warped supergravity, that is, 5d supergravity with a $U(1)$ subgroup of the $SU(2)$ R-symmetry group gauged by the graviphoton \cite{ceag}.
The 5d action can be obtained from the flat one in \eref{5da} by replacing all the derivatives acting on the gravitino by
\beq
D_\alpha \psi_\beta \to
  D_\alpha \psi_\beta  
+{i \over 2} k \epsilon(x_5) \Gamma_\alpha \psi_\beta 
+  {i\sqrt{6} \over 2} k \epsilon(x_5) \ca_\alpha \psi_\beta  \, .
\eeq
Analogous replacement should be done for $D_\alpha \epsilon$ in the gravitino transformation laws. Besides, the 5d bulk action\erefn{5da} should  be supplemented by
\beq
\ll{e.5dwa}
\cl = 6 M_5^3 k^2e_5     \, .
\eeq
Hence the gauging implies the presence of a negative cosmological constant in the bulk and so the gravitational background solution is ${\rm AdS}_5$. On the orbifold, the presence of the step function $\eps(x_5)$ induces 
additional variations proportional to the delta function,
\beq
\label{e.wjsv}
\delta \cl = M_5^3 e_4 \delta(x_5) k 
\left [ 3 i \ov{\psi_\mu}\gamma^\mu \gamma^5 \epsilon
- \sqrt{6} i \ov{\psi_\mu}\gamma^{\mu\nu} \gamma^5 \epsilon \ca_\nu  \right ]
 (\delta(x_5) - \delta(x_5-\pi R)) \, .
\eeq
Canceling the first term  requires  the presence of the  brane tension \cite{ghpo},
\beq
\cl =  - 6 M_5^3 e_5  k (\delta(x_5) - \delta(x_5-\pi R)) \, .
\eeq
and so  the gravitational background in this set-up is precisely that of the Randall-Sundrum model \cite{rasu}. In absence of brane matter the second term vanishes.
 When brane chiral multiplets are present  the current $j_\mu^0$ in \eref{hfs} acts  as a source for the  graviphoton so that it has a jump at the brane, 
 $\ca_{\mu} \sim \eps(x_5) {1 \over 2 M_5^3} j_\mu^0$. 
Moreover, the Noether term \erefn{nt} is a  source for the negative parity gravitino.  By equations of motion it behaves as  
$\gamma^{\mu\nu} \psi_\nu^-  
\sim  {1 \over 2 \rt}\eps(x_5) \gamma^\rho \gamma^\mu  \psi_Q \pa_\rho Q_0$  near the brane. In the flat case $k \to 0$ these subtleties in boundary conditions do not affect the Noether construction at the level of two-fermion terms. But for $k \neq 0$ these boundaries conditions imply that the second term in \eref{wjsv} is non-vanishing.\footnote{A similar treatment of the boundary conditions in supersymmetric variations was also employed in  refs. \cite{babe,babe2}.} It turns out that the necessary modification that cancels this term consists in multiplying the brane action by a $|Q_0|^2$ dependent factor, 
\beq
\omega_0(|Q_0|^2) = {1  \over  1 - {k |Q_0|^2 \over 3 M_5^3}}
\eeq
 The brane-world action up to four-fermion terms reads
\bea  &
\ll{e.wba}
\cl = \cl_{\rm bulk} 
(\cf_{\mu 5} \to \cf_{\mu  5} + {1 \over M_5^3} \delta(x_5) j_\mu^0 ) 
+ \delta(x_5) \cl_{\rm brane}  \, ,
\nl \ds 
j_\mu^0 =  {i \over \sqrt 6} \omega_0
\left (Q_0^\dagger \pa_\mu Q_0 -\pa_\mu Q_0^\dagger Q_0  \right )
- {1 \over 2\sqrt 6} \omega_0^2 \psi_Q \gamma_\mu \psi_Q  \, ,
\nl \ds 
\cl_{\rm brane}  =  
- 6 M_5^3 k e_5 
+e_4 \omega_0^2 
 \left [
 {1 \over 2} \pa_\mu Q_0^\dagger \pa^\mu Q_0 
\bnl  \ds
+ {i \over 2} \ov \psi_Q \gamma^\mu D_\mu \psi_Q 
  - {1 \over \rt} e_4 \delta(x_5)  
\ov \psi_Q \gamma^\mu  \gamma^\rho \pa_\rho Q_0 \psi_\mu + \hc \, .
\right ] 
\eea
One should also insert the factor $\omega_0$ into the transformation of the graviphoton in \eref{gt}. 
The action for the matter on the brane at $x_5 =\pi R$ is analogous with 
$\delta(x_5) \to \delta(x_5-\pi R)$, $k \to -k$, $Q_0\to Q_\pi$,  $\psi_Q\to \psi_{Q_\pi}$.  

One can further extend the model to include  a non-trivial superpotential 
$W_0(Q_0)$ for the brane multiplet. The treatment of the boundary conditions is then much more involved  and 
in this paper we only discuss some limiting cases.  
 Howver this discussion is  not necessary for our  computation of one-loop corrections and 
so we  shift it to \aref{bsp}.

\section{Four-dimensional  effective supergravity}
\label{s.elt}

We move to discussing  the form of the 4d effective supergravity that describes the light degrees of freedom  
(those with masses below the compactification scale) of the 5d theory compactified on the background
\beq
\ll{e.brs}
ds^2 = a^2(x_5) g_{\mu\nu} dx^\mu dx^\nu - \phi^2 dx_5^2 
\qquad  a(x_5) = e^{- k \phi x_5} \, .
\eeq
The limit $k \to 0$ corresponds to flat compactification. 
The bosonic degrees of freedom are the 4d metric $g_{\mu\nu}$, the radion $\phi \equiv \sqrt{g_{55}}$, the fifth component of the graviphoton $\ca_5$ and the scalars on the two branes $Q_0$ and $Q_\pi$. The \kahler potential of the 4d supergravity can be found by matching with  the kinetic terms for those fields. 

The kinetic terms for the metric component are obtained by inserting the background \erefn{brs}  into the 5d Einstein-Hilbert action,  
\beq
\ll{e.tmp1}
\cl = \sqrt{-g} M_p^2 \left[ {1 \over 2} {1 - a_\pi^2 \over 2 \kr}  R(g) + {3 \over 2} \kr \,  a_\pi^2 (\pa_\mu \phi)^2  \right] \, ,
\eeq
where $a_\pi = e^{-\kr \phi}$ and $M_p^2 = 2 \pi R M_5^3$. 
To go the Einstein basis we need to perform the Weyl rescaling  
$g_{\mu\nu} \to {1 \over f_E(\phi)}g_{\mu\nu}^{(E)}$ with 
$f_E(\phi) = {1 - a_\pi^2 \over 2 \kr}$. Then the kinetic terms become
\beq
\ll{e.kg}
\cl = \sqrt{-g^{{}_{(E)}}} M_p^2 \left[ {1 \over 2}   R(g) 
+ {3 \over 4 f_E(\phi)^2 }   a_\pi^2 (\pa_\mu \phi)^2  \right] \, .
\eeq
The kinetic terms in \eref{wba} yield
\beq
\label{e.kq}
\cl =\sqrt{-g^{{}_{(E)}}} {1 \over f_E(\phi)} 
\left [\omega_0^2 \pa_\mu Q_0^\dagger \pa^\mu Q_0
 + a_\pi^2 \, \omega_\pi^2 \, \pa_\mu Q_\pi^\dagger \pa^\mu Q_\pi \right] \, .
\eeq
More care is needed to derive kinetic terms of $\ca_5$. To do this consistently we need to integrate out the negative parity components of the graviphoton $\ca_\mu$. The relevant part of the 5d action is 
\beq
\cl_{5d} = {1 \over 2 \phi}  a^2(x_5) \sqrt{-g} M_5^3 \left (
\pa_5 \ca_\mu - \pa_\mu \ca_5
 + {1 \over M_5^3} \delta(x_5) j_\mu^0 
+ {1 \over M_5^3} \delta(x_5-\pi R) j_\mu^{\pi} \right )^2  \, .
\eeq
The solution to the graviphoton equations of motion is  
$\ca_\mu = \pa_\mu \ca_5 x_5 + {1 \over 2 k} \eps (x_5) C_\mu a^{-2}(x_5) + \eps(x_5) D_\mu$.  
The boundary conditions $\ca_\mu(0) = j_\mu^0/2$, $\ca_\mu(\pi R) = -j_\mu^\pi/2$ determine the integration constants and we find  
$C_\mu = -a_\pi^2 {\pa_\mu \ca_5 + (j_\mu^0+j_\mu^\pi)/M_p^2 \over f_E(\phi)}$. 
Inserting this solution back into the 5d action and integrating over $x_5$ yields
\beq
\ll{e.ka}
\cl_4 =  {1 \over 2} M_p^2 \sqrt{-g^{{}_{(E)}}} 
{a_\pi^{2} \over f_E(\phi)^2}
 \left (\pa_\mu \ca_5 + {1 \over M_p^2} (j_\mu^0+j_\mu^\pi) \right )^2  \, .
\eeq
Note that the $\delta^2$ terms has canceled. 
The \kahler potential that reproduces the kinetic terms \erefn{kg}, \erefn{kq} and \erefn{ka} is given by $K = - 3 \log \Omega$ where 
\bea&
\ll{e.krsm} 
\ds 
\Omega = {1  - e^{- \kr (T+\ov T)} \over 2 \kr} 
 - {1 \over 3 M_p^2} |Q_0|^2 - {1 \over 3 M_p^2} e^{- \kr (T+\ov T)} |Q_\pi|^2 \, ,
\nl \ds 
{\rm Re\, } T = \phi 
 - {1 \over 2 \kr} \log\left(1 - {2 \kr \over 3 M_p^2} |Q_0|^2 \right )
 + {1 \over 2 \kr} \log\left(1 + {2 \kr \over 3 M_p^2}  |Q_\pi|^2 \right ) \, ,
\nl \ds
 {\rm Im \,} T = i \sqrt{2 \over 3} \ca_5 \, .
\nn
\eea 
One can check that also the remaining interaction terms in the brane-world action \erefn{wba} fit the general structure of 4d supergravity \cite{crjusc} with the \kahler potential of \eref{krsm}. 
Generalization to an arbitrary number of brane matter multiplets with general 
kinetic terms  is straightforward. It amounts to replacing $|Q_i|^2$ with 
arbitrary real functions $\Omega_i(Q_i^n)$  
\bea&
\ll{e.wtkg} 
\ds 
\Omega = {1  - e^{- \kr (T+\ov T)} \over 2 \kr} 
 - {1 \over 3 M_p^2}\Omega_0(Q_0^n)
 - {1 \over 3 M_p^2} e^{- \kr (T+\ov T)} \Omega_\pi(Q_\pi^m)  \, ,
\nl \ds 
 {\rm Re\, }T = \phi 
-{1 \over 2 \kr} \log\left(1 - {2 \kr \over 3 M_p^2} \Omega_0(Q_0^n) \right )
+{1 \over 2 \kr}\log\left(1 + {2 \kr \over 3 M_p^2}\Omega_\pi(Q_\pi^m) \right ) \, .
\nn
\eea 
In the flat limit $k\to 0$ we recover the well-known no-scale structure,
\bea&
\ll{e.kfl} \ds 
\Omega = {T + \ov T \over 2}  - {1 \over 3 M_p^2} \Omega_0(Q_0^n) - {1 \over 3 M_p^2} \Omega_\pi(Q_\pi^m) \, ,
\nl \ds 
 T = \phi + {1 \over 3 M_p^2}  \Omega_0(Q_0^n) + {1 \over 3 M_p^2}\Omega_\pi(Q_\pi^m)   + i \sqrt{2 \over 3} \ca_5  \, .
\eea 

Furthermore, in the presence of brane superpotential 
$W_0(Q_0^i)$ and $W_\pi(Q_\pi^i)$ the  superpotential of the effective 
4d \sugra reads
\beq
\ll{e.wtsp}
W = W_0(Q_0^i)  + e^{- 3 \kr T} W_\pi(Q_\pi^i) \, .
\eeq  
The \kahler potential derived here is the same function of $T$ and $Q$ as the one in ref. \cite{grrasc} 
(note that we use the definition of $\Omega$ that differs by a factor $-1/(3 M_p^2)$ from that of ref \cite{grrasc}).
However the definition of the modulus $T$ in terms of the 5d degrees of freedom is different (in our formulation it is also a function  the brane matter fields). Of course, at tree-level the physics 
(like moduli stabilization, transmission of supersymmetry breaking) is the same in both formalisms.
In particular  the \kahler potential in \eref{wtkg} implies no tree-level mediation of supersymmetry breaking through the bulk 
(although for $k \neq 0$ it is not of the no-scale form).

\section{One-loop corrections to the \kahler potential} 
\ll{s.olc}

We now use our on-shell formulation of the 5d theory to compute one-loop corrections  to  the tree-level \kahler potential. From the point of view of the 4d effective theory no symmetry protects the particular structure of $\Omega$ in  \eref{wtkg}. We expect that $\Omega_{1\;\rm loop} = \Omega + \Delta \Omega$ and that $ \Delta \Omega$  includes couplings other than those in \eref{wtkg}, for example higher powers of $e^{- \kr (T+\ov T)}$ or contact terms between $Q_0$ and $Q_\pi$. These new terms will lead to mediation of supersymmetry breaking.  

We  first compute the one-loop effective action in the full 5d theory and then match to 4d effective supergravity with a \kahler potential 
$\Omega + \Delta \Omega$. 
The computation involves regularization of divergent expressions so we first discuss the most general structure of the counterterms in the \kahler potential.  Since $\Omega = e^{-K/3}$ is the coefficient of the  Einstein-Hilbert term in the supergravity conformal frame, 
\beq
\ll{e.ktcf}
\cl_{C} = \sqrt{-g^{{}_{C}}} M_p^2 \left [
{1 \over 2} \Omega R 
- 3 \Omega_{\ov m n}\pa_\mu z_{\ov m}^\dagger \pa_\mu z_n 
  - {3 \over 4 \Omega} (\Omega_{\ov m }\pa_\mu z_{\ov m}^\dagger -   \Omega_{n}  \pa_\mu z_n)^2 + \dots \right ] \, ,
\eeq
the possible counterterms are constrained by the most general form of the Einstein-Hilbert terms consistent with 5d general coordinate invariance and locality, 
\beq
\cl = C_B \sqrt{-g_5} R_5  + C_0(Q_0) \delta(x_5) \sqrt{-g_4} R_4   + C_\pi(Q_\pi) \delta(x_5-\pi R) \sqrt{-g_4} R_4 \, .
\eeq  
After compactification on the warped background \eref{brs}
and Weyl rescaling to the conformal frame 
$g_{\mu\nu} \to  f_C g_{\mu\nu}^{{}_{C}}$ this becomes
\beq
\cl = \left [C_B {1 - a_\pi^2  \over 2 \kr} 
+ C_0(Q_0) +  C_\pi(Q_\pi) a_\pi^2 \right ] f_C \sqrt{-g^{{}_{C}}} R  \, .
\eeq
In our case $f_C =  1 - {2 \kr \over 3 M_p^2} \Omega_0(Q_0)$.  
Using \eref{wtkg} we express $\phi$ in $a_\pi = e^{- \kr \phi}$ by $T$ and $Q$ and we obtain
\bea  & \ds
\cl = \left [  
\left ({1 \over 2 \kr } C_B + C_0(Q_0) \right)  
\left ( 1 - {2 \kr \over 3 M_p^2} \Omega_0(Q_0)\right)  
\bnl
+ e^{-\kr(T+\ov T)}
\left (-{1 \over 2 \kr } C_B + C_\pi(Q_\pi) \right)
 \left ( 1 + {2 \kr \over 3 M_p^2} \Omega_\pi(Q_\pi)\right) 
\right] \sqrt{-g^{{}_{C}}} R  \, .
\eea
We see that the coefficient of the Einstein-Hilbert term is of the same form as the \kahler potential \erefn{wtkg}. We are thus guaranteed that all  divergences we encounter in the one-loop computation can be absorbed by renormalization of the parameters in the tree-level \kahler potential \erefn{wtkg}.
 In particular, these divergences are not relevant for the questions of  supersymmetry breaking mediation through the bulk. 
On  the other hand, any couplings in $\Delta \Omega$ that are different than those in the tree-level \kahler potential correspond necessarily to  non-local operators in 5d theory  and therefore they should be finite and UV insensitive.

Reference \cite{grrasc} derives a very useful expression for  $\Delta \Omega$, 
\beq
\ll{e.dor}
\Delta \Omega  \sim \intdk \sum_n {1 \over k^2} \log(k^2 + m_n^2) \, . 
\eeq  
 Before we compute  $\Delta \Omega$ in our set-up we first present a simple derivation of \eref{dor}.
In order to compute $\dO$ it is sufficient to compute corrections to the Einstein-Hilbert term in the conformal frame and compare the result with \eref{ktcf}. Quite generally, a field of spin $j$ and mass $m$ contributes to one-loop renormalization of the Einstein-Hilbert term (in dimensional regularization):
\beq
\ll{e.drg}
\Delta \cl_{j} = n_j {\Gamma(1-d/2) m^{d-2} \over (4 \pi)^{d/2}} \sqrt{-g} R \, ,
\eeq
where $n_0 = 0$ for a conformally coupled scalar,
 $n_{1/2} = 1/6$ for a Dirac fermion,  $n_1 = -1/3$ for a gauge boson,
and $n_{3/2} + n_2 = 0$ for a summed contribution of a Dirac gravitino and a graviton.  Specializing to the case of 5d sugra, a 5d  hypermultiplet contains one Dirac fermion, a 5d vector multiplet - one Dirac fermion  and one gauge boson, while the gravity multiplet contains one gauge boson, one  Dirac gravitino and one graviton  at each KK level. Summing all these contributions we find:
 \beq
\ll{e.dog}
\dO = {1 \over 3} (- 2 - N_V + N_H) {\Gamma(1-d/2)\over M_p^2 (4 \pi)^{d/2}} \sum_n m_n^{d-2}  \, .
\eeq   
where $N_V$ and $N_H$ is the number of vector multiplets and hypermultiplets, respectively, and  $m_n$ are the masses of the KK modes in the conformal frame as a function of background values of $T$ and $Q$.   
For $N_V = N_H = 0$ this formula is equivalent to that in  ref. \cite{grrasc} with the momentum integral evaluated using dimensional regularization.

We now apply the general formula \erefn{dog} to the model considered in this paper. In the frame set by $g_{\mu\nu}$ in \eref{brs} the KK spectrum is given by positive roots of the equation:
\beq
\ll{e.kksw}
J_1\left ({m_n \over k} \right )Y_1 \left ({m_n \over k a_\pi} \right ) 
- Y_1\left ({m_n \over k} \right )J_1 \left ({m_n \over k a_\pi} \right ) = 0 \, .
\eeq
Note that in our set-up the KK spectrum $m_n$ is a function of the field $\phi$ only and is not modified by the presence of brane matter fields $Q$. 
Going to the conformal frame $m_n \to m_n f_C^{1/2}$, 
so $\dO$ picks up an additional multiplicative factor $f_C^{d/2 -1}$. 

Using the standard tools \cite{popo} we convert the sum over KK modes into a contour integral. The divergent part is of the form $f_C(C_1 + a_\pi^2 C_2)$ and can be absorbed into renormalization of the tree-level \kahler potential. The remaining finite part is given by
\bea & \ds
\label{e.dow}
\dO =  {4 \over 3 M_p^2 (4 \pi)^2} 
\left (1 - {2 \kr \over 3 M_p^2} \Omega_0(Q_0) \right )
 k^2  a_{\pi}^2\int_0^\infty dy y 
\log \left (1 - { I_1 (y a_{\pi})K_1 (y)  \over K_1 (y a_{\pi})I_1 (y)} 
\right ) 
\nl \ds 
a_\pi^2  =  e^{-\kr (T+\ov T)}
{ 1 + {2 \kr \over 3 M_p^2}\Omega_\pi (Q_\pi)   \over 
1 - {2 \kr \over 3 M_p^2} \Omega_0(Q_0)} \, .
\eea 
The \kahler potential $\Omega + \dO$ contains all information about the contact terms between the hidden and observable sectors.
In the limit of large warping, $a_\pi \to 0$  we can approximate $\dO$ by:
\beq
\label{e.dowl}
\dO \approx  -{4 c k^2 \over 3 M_p^2 (4 \pi)^2} e^{- 2 \kr (T+\ov T)}
{ \left (1 + {2 \kr \over 3 M_p^2}\Omega_\pi (Q_\pi) \right )^2 
 \over 1 - {2 \kr \over 3 M_p^2} \Omega_0(Q_0)}
 +\co(a_\pi^6\log a_\pi) \, ,
\eeq
where $c= \int_0^\infty dy y^3 {K_1(y) \over 2 I_1(y)} \approx 1.165$.  
On the other hand in the flat limit $k \to 0$ we find
\beq
\ll{e.dof}
\Delta \Omega  = - {16 \zeta(3) \over 3 (4 \pi)^2} {1 \over (2 \pi R M_p)^2} 
{1 \over \left( T + \ov T - {2 \over 3 M_p^2}\Omega_0(Q_0) - {2 \over 3 M_p^2}\Omega_\pi(Q_\pi) \right)^2}  \, .
\eeq 

\section{On ambiguity in one-loop \kahler potential}
\ll{s.akp}

We now compare the result of our computation to the previous works on the subject \cite{ghri,rascst,bugago}. 
For simplicity, we restrict to the flat limit $k \to 0$. 
Expanding the $\dO$ in \eref{dof} in powers of $\Omega_i$ we obtain 
\bea & 
\ll{e.dofe}
\Delta \Omega = - {16 \zeta(3) \over 3 (4 \pi)^2} {1 \over (2 \pi R M_p)^2} 
\left [ 
{1 \over(T + \ov T)^2} 
 + {4 \over 3 M_p^2 (T + \ov T)^3} ( \Omega_0(Q_0)+\Omega_\pi(Q_\pi) )
\bnl \ds
 + {4 \over 3 M_p^4 (T + \ov T)^4} (\Omega_0(Q_0)+\Omega_\pi(Q_\pi))^2 + \dots 
\right ] \, .
\eea
The first term describes the Casimir energy \cite{popo}, the second corresponds to radion mediation \cite{ghri} and the last one to brane-to-brane mediation of supersymmetry breaking \cite{rascst}. At this order, all the terms in \eref{dofe} are the same as those derived in the literature.  However the full formula \eref{dof} is clearly different than that in ref. \cite{rascst} and the  discrepancy enters at the cubic order in $\Omega_i$. As  long as the brane sources are perturbative, the physical consequences of both formulations (summarized in  \eref{dofe}) are the same.
 However when the brane sources are large (for $\Omega_i > 2 \pi R \phi M_5^3$) conclusions derived  in both formalisms may be completely different.
 In particular, from \eref{dof} it is evident that a constant terms in the boundary \kahler potential, $\Omega_i = L + \dots$, is equivalent to shifting $T$ by a constant and therefore has no physical significance.
 This is certainly different than in ref. \cite{rascst} where a large value of $L$ was needed for obtaining positive soft mass terms.

 The origin of this incompatibility can be traced to the different formulation of the 5d brane-world theory.
The technical issue that affects the one-loop computation is the fact that
 in the off-shell formulation   $\Omega_0$ and $\Omega_\pi$ multiply brane kinetic terms of the gravity multiplet. These couplings remain after integrating out the auxiliary fields. On the other hand, in our purely on-shell Noether formulation such terms  are absent.   
In order to understand this difference better, below we discuss 
supersymmetrization of a model with a brane Einstein-Hilbert term by means of the  Noether procedure. For simplicity we restrict to the case where no brane matter is present. Thus we start with the brane lagrangian of the form
\beq
\ll{e.dgp1}
\cl = e_4 M_5^3 L \delta(x_5) \left [ 
{1 \over 2} R_4 - i \ov \psi_\mu^+ \gamma^{\mu\nu\rho} D_\nu \psi_\rho^+
 \right]  \, .
\eeq
In ordinary 4d \sugra this lagrangian would be supersymmetric up to  four-fermion terms. But here $\delta \psi_\mu \sim \cf_{\mu 5} \epsilon$ and so the variation of \eref{dgp1} is non-zero already at the two-fermion level. To cancel it one has to add new zero- and two-fermion terms to \eref{dgp1} as well as modify the supersymmetry transformation of the gravitino by terms proportional to $L \delta$. 
However once one arrives at a lagrangian in which all variations of order $L\delta$ cancel one finds that there are  $L^2 \delta^2$ variations that do not cancel. Therefore the Noether procedure must be continued and new singular terms of order  $L^2 \delta^2$ have to be added to \eref{dgp1} to make the lagrangian supersymmetric at this order\footnote{In the following we ignore the mathematical subtleties involved in multiplication of distribution  and manipulate $\delta$'s as if they were ordinary c-numbers}.
The story does not end at order $\delta^2$. In order to maintain supersymmetry singular terms with higher and higher powers of $\delta$ are needed. 
However one can notice that there is a certain pattern emerging. It turns out that 
all the terms obtained by the Noether procedure can be obtained from the bulk action 
\beq
\cl  =  e_5 M_5^3 \left [ 
{1 \over 2} R_5 - i \ov \psi_\alpha \Gamma^{\alpha\beta\gamma}
 D_\beta \psi_\gamma - {1 \over 4} \cf_{\alpha\beta} \cf^{\alpha\beta}+\dots \right]  
\eeq
by a formal, singular redefinition of the  $\phi \equiv e_{\dot 5}^5$
 component of the 5d vielbein,   
\beq
\ll{e.dgprl}
\phi \to \phi + L \delta (x_5) \, .
\eeq
In addition, one should assume that only positive parity fields multiplied by $\delta(x_5)$ survive in the brane action. 
 
For example by Noether procedure we get a series of graviphoton brane kinetic terms
\bea & \ds
\ll{e.dgp2}
\cl = {1 \over 2 \phi} M_5^3 e_4 
\left [ {L \over \phi} \delta(x_5)
 - {L^2\over \phi^2} \delta(x_5)^2
 +  {L^3 \over \phi^3} \delta(x_5)^2 + \dots \right ] \cf_{\mu 5}^2 
\nl \ds
 = {1 \over 2} \left({1 \over \phi + L \delta(x_5)}- {1 \over \phi} \right )
 M_5^3 e_4 \cf_{\mu 5}^2 \, .
\eea
 One can argue that the extremely singular term like the one in \eref{dgp2} is indeed needed for supersymmetry. Indeed, the graviton and gravitino KK spectrum with brane kinetic terms of \eref{dgp1} is given by solutions of 
\beq
\ll{e.dgps}
\tan (\phi \pi R m_n) = -{1 \over 2} m_n L  \, .
\eeq   
For the graviphoton the equation of motion reads
$\pa_5 \left ( {\pa_5 \ca_\mu \over \phi + L \delta(x_5)} \right )
+ \phi m_n^2 \ca_\mu  = 0
$
 To cancel all $\delta$'s we should arrange that $\pa_5 \ca_\mu$ behaves as $ \sim \phi + L \delta(x_5)$ near the boundary. Matching this boundary condition with the bulk solution of the equations of motion we get precisely the quantization condition \eref{dgps}. Getting this spectrum for the graviphoton would be impossible without brane $\cf^2$ term or with any decent non-singular $\cf^2\delta$ term. We conclude that for supersymmetrizing a 5d model with a boundary Einstein-Hilbert term it is indeed necessary  to include an infinite series of singular $\delta^n$ terms  in the action.   
 
We can now infer the relation between the brane-world action obtained by integrating out auxiliary fields in the off-shell formulation  and the one obtained by our Noether procedure.  The two are connected by a singular change of variables
\beq
\ll{e.dgprg}
\phi \to \phi -  {1 \over 3 M_5^3} \Omega_0(Q_0) \delta (x_5) 
-  {1 \over 3 M_5^3} \Omega_\pi(Q_\pi) \delta (x_5-\pi R) \, .
\eeq
If the two formalisms  are in fact equivalent up to a change of variables why the computation of  loop corrections yields different results? The difference can be traced to ambiguity of defining the behavior of bulk fields near the $\delta$ sources. In the above example, after the redefinition \erefn{dgprl} we kept  only  positive $Z_2$ parity fields in the brane action. But
once we switch on a source of order $L \delta$, by equations of motions  the  negative $Z_2$ parity fields behave like $\sim L \eps(x_5)$ near the boundary. We are then allowed to keep also boundary terms involving $Z_2$ odd fields,
 $\cl \sim (\psi^-)^2 \delta(x_5)$, provided we define 
the distribution $\delta(x_5) \epsilon^2(x_5)$ to be non-vanishing. 
 Such terms  affect the KK spectrum at the cubic order in $L$ and, by \eref{dog}, also the one-loop \kahler potential at higher order in  $L$.

Concluding, the change of variable \eref{dgprg} defines in fact a class of brane-world actions, depending on what regularization scheme we adopt to resolve the brane singularity. Physical predictions within this class of theories may differ at the third order in brane sources.   As long as the brane sources are perturbative, the relevant physical quantities (e.g. soft mass terms) derived  in both formulations  are the same. However, if the brane sources are large (in the above example, if $L$ is bigger that the compactification length $2 \pi R \phi$) the low-energy observables may depend on how the brane singularity is regularized.      

\section{Conclusions}

In this paper we used the Noether method to construct 5d on-shell gauged supergravity coupled to chiral matter multiplets
 on the branes. This turned out to lead to a slightly different set-up than that derived from  the more commonly used off-shell formulation. Certain singularities that appear after integrating out the auxiliary fields are absent in the purely on-shell  Noether formulation. This is due to a different choice of the fundamental degrees of freedom in the 5d theory. 

Furthermore, we showed that our on-shell set-up  allows for a simple computation of one-loop corrections.
Comparison of our results with previous works showed an ambiguity in computation of the one-loop \kahler potential.
 This ambiguity is associated with  arbitrariness in resolving the singularities associated with infinitely thin, delta-type branes. As long as the brane sources are small (the reference scale being $M_5^3 2 \pi R \phi$) this ambiguity has negligible effects on the low-energy physics. However in certain 5d models large brane sources are essential. 
One well-known example is the Dvali-Gabadadze-Porratti \cite{dvgapo} model in which gravity is localized on a 4d brane in a  semi-infinite flat extra dimension. We conclude that there is a whole class of supersymmetric completions of the DGP model that yield different low-energy predictions. Another example are set-ups with a gravitino brane mass term, 
$\cl \sim W \delta(x_5) \psi_\mu^T \gamma^{\mu\nu} \psi_\nu + \hc$.
The limit $W \to \infty$ is sometimes considered as being equivalent to the set-up with supersymmetry broken by boundary conditions. In such limit there  is also a continuous family of regularizations that results in different physics at low energies.  

We expect that the Noether method can be easily carried over to other brane-world models, for example to  6d supergravity with matter on a brane of  co-dimension two. This offers an opportunity to construct more general brane-world actions and study its low-energy phenomenology without a necessity of going through the off-shell calculus.

\vspace{5mm}
\noindent{\large\bf Acknowledgments}
\vspace{5mm}

I would like thank Stefan Pokorski for numerous discussions and collaboration on some stages of this project. I am also grateful to Riccardo Rattazzi and Claudio Scrucca for useful discussions.

The author  was partially supported by thr Polish KBN grant 2 P03B 129 24 for years 2003-2005
and by the EC Contract MRTN-CT-2004-503369 - network "The Quest for Unification: Theory Confronts Experiment" (2004-2008). The stay at DESY is possible due to Research Fellowship  granted by  Alexander von Humboldt Foundation.

\renewcommand{\thesection}{Appendix~\Alph{section}}
\renewcommand{\theequation}{\Alph{section}.\arabic{equation}}
\setcounter{section}{0}
\setcounter{equation}{0}
\section{Notation and conventions}
\label{a.nc}
We use the mostly minus metric signature $(+,-,-,-)$. The index conventions are the following: 5d Einstein indices $\alpha,\beta,\gamma \dots = 0\dots 3,5$,  5d Lorentz indices $a,b,c \dots = 0\dots 3,5$, 4d Einstein indices $\mu,\nu,\rho \dots = 0\dots 3$,  4d Lorentz indices $m,n,\dots = 0\dots 3$. 

The 5d vielbein is denoted by $e_\alpha^a$ and satisfies  $e_\alpha^a e_\beta^b \eta_{ab} = g_{\alpha\beta}$. $e_5$ is the determinant of the 5d vielbein, while by $e_4$ we denote the determinant of the 4d vielbein induced at the boundary. Similarly, $R_5$ denotes the 5d Ricci scalar, while  $R_4$ denotes the Ricci scalar constructed from the 4d vielbein induced at the boundary.  The inverse vielbein  $e^\alpha_a$ satisfies $e^\alpha_a e_\beta^a = \delta_\alpha^\beta$.  

5d gamma matrices are denoted as $\Gamma^{a}$ while 4d gamma matrices are denoted as $\gamma^{m}$. They satisfy $\{ \Gamma^{a}, \Gamma^{b} \} =  2 \eta^{ab}$ and 
$\{ \gamma^{m}, \gamma^{n} \} = 2 \eta^{m n}$. 
The connection between the two sets is given by $\Gamma^m = \gamma^m$, $\Gamma^5 = i\gamma^5$. Furthermore 
$\Gamma^\alpha = e^{\alpha}_a \Gamma^a$, 
$\gamma^\mu = e^{\mu}_m \gamma^m$. 
The convention for $\gamma^5$ is 
$\gamma^5 = {\rm diag} (-1,-1,1,1) $
 and the chirality projection operators are $P_L = (1 - \gamma^5)/2$, $P_R = (1 + \gamma^5)/2$.   All the fermions in 5d and 4d are in four-component Dirac notation (we don't use symplectic Majorana spinors). 
The 4d charge conjugation matrix $C = i \gamma^0 \gamma^2 \gamma^5$ satisfies  
$C^{-1} = C^T = C^\dagger = -C$,   
$ C \gamma^{m} C^{-1} = (\gamma^m)^T$.

\section{Brane action with a superpotential}
\label{a.bsp}

One can extend the set-up studied in \sref{bwl} to include  a non-trivial superpotential $W_0(Q_0)$ for the brane multiplet.
 We start with the case of flat supergravity in the bulk. In addition to those of \eref{gsl}, the terms present in the globally supersymmetric limit are the following
\beq
\cl_{3} = e_4 \delta (x_5) \left [
 - \left | {\pa W_0 \over \pa Q_0} \right |^2  
+ {1 \over 2} {\pa^2 W_0 \over \pa Q_0^2} \psi_Q^T C \psi_Q 
- {1 \over 2} {\pa^2\ov{W_0} \over \pa Q_0^{\dagger 2}} \ov \psi_Q C {\ov \psi_Q}^T \right ] \, .
\eeq 
The supersymmetry  transformation of the chiral fermion is supplemented by 
$\delta \psi_Q = {1 \over \sqrt 2}{\pa \ov{W_0} \over \pa Q_0^\dagger} C \ov{\eps}^T$.  In the presence of the superpotential the matter supercurrent is modified. The gravitino couples additionally as 
\beq
\ll{e.ntw}
\cl_{4} = -{i \over \sqrt 2}e_4 
\delta (x_5){\pa W_0 \over \pa Q_0} \psi_Q^T C \gamma^\mu \psi_{\mu} + \hc \, .
\eeq 
Besides, up to four-fermion terms local supersymmetry requires one more term
\beq
\label{e.bm2}
\cl_{5} = e_4 \delta(x_5) \left[
- {1 \over 2} W_0 {\ov\psi_\mu}  \gamma^{\mu\nu} C \ov{\psi_\nu}^T + \hc 
 \right] \, .
\eeq
Furthermore one modifies the trasformation law of $\psi_5$ by
\beq
\delta \psi_5 = - \delta(x_5) W_0  C \ov{\eps}^T \, . 
\eeq
The action on the other brane is analogous with $\delta(x_5) \to \delta(x_5 - \pi R)$, $W_0(Q_0) \to W_\pi(Q_\pi)$. 
Again, no singular terms arise in this construction.
One important comment is in order here. 
The gravitino brane mass term acts as a source for negative $Z_2$ parity gravitino so that 
 $M_5^3 \psi_\mu^- \sim \eps(x_5) W_0 C \ov{\psi_{\mu}^+}^T$ near the boundary. Thus, in general, $\psi_\mu^-$ can also couple to the brane matter. Therefore we have to reconsider the question which combination of the two bulk gravitinos should  couple to the matter supercurrent in \eref{nt} and \eref{ntw}. It turns out that the answer depends on how the delta singularity is regularized. But  whatever regularization we choose there is always one combination of the two bulk gravitinos $\eps(x_5) \psi_\mu^+ \sin \theta + C\ov{\psi_{\mu}^-}^T \cos \theta$ 
that vanishes at the brane in the limit when the regulator is removed. Then the orthogonal combination $
\psi_\mu =   \psi_\mu^+ \cos \theta  
- \eps(x_5) C\ov{\psi_{\mu}^-}^T \cos \theta $  couples to the brane matter. The angle of rotation is given by 
$\theta = W_0/(2 M_5^3) + \co (W_0^3)$, where the higher order terms in $W_0$
 are regularization dependent. Coupling this combination  $\psi_\mu$  to   the matter supercurrent yields also correct (consistent with the general supergravity action \cite{crjusc}) couplings in the 4d effective action, up to $\co (W_0^3)$ terms. But another conclusion from this discussion is that the low-energy 4d supergravity is regularization independent only up to terms cubic in the brane superpotential $W_0$.    

A careful treatment of the boundary conditions for gravitinos is even more important when warped supergravity ($k \neq 0$) is present in the bulk.
 Ref. \cite{babe} discussed this problem for the case when brane matter is absent and $W_0$ is a constant. It was found that the brane action has to be modified already at the purely bosonic level. The brane tension term is given  by:
\beq
\ll{e.ww}
\cl = - 6 M_5^3 k e_4 \delta(x_5) {1 - {W_0^2 \over 4 M_5^6} \over 1 + {W_0^2 \over 4 M_5^6}} \approx 
 - 6 M_5^3 k e_4 \delta(x_5) ( 1 - {W_0^2 \over 2 M_5^6}) + \co(W_0^4) \, . 
\eeq
There is a similar term on the other brane with $k \to - k$, $W_0 \to W_\pi$.  Therefore, in the presence of gravitino brane mass terms the Randall-Sundrum tuning between the bulk cosmological constant and the brane tension is lost. The background solution is then of  the Randall-Karch type \cite{raka} with   
${\rm AdS}_4$ symmetry of the 4d spacetime. Note that is consistent with what we obtain in the low energy 4d supergravity description. With the \kahler potential \eref{wtkg} and the superpotential \eref{wtsp} we obtain the scalar potential 
$V = {6 \kr \over f_E(\phi)^2 M_p^2} (|W_0|^2 - e^{- 4\kr \phi}|W_\pi|^2)$
 which is  what follows from the warped compactification with the brane tension of \eref{ww}. 

The case when warped supergravity, brane matter and brane superpotentials are present simultaneously is technically more involved and will be left for future studies. 



\begin{thebibliography}{99}
\bibitem{rasu0}
L.~Randall and R.~Sundrum,
Nucl.\ Phys.\ B {\bf 557} (1999) 79
[arXiv:hep-th/9810155].

\bibitem{gilumu}
G.~F.~Giudice, M.~A.~Luty, H.~Murayama and R.~Rattazzi,
JHEP {\bf 9812} (1998) 027
[arXiv:hep-ph/9810442].

\bibitem{ghri}
T.~Gherghetta and A.~Riotto,
Nucl.\ Phys.\ B {\bf 623} (2002) 97
[arXiv:hep-th/0110022].

\bibitem{bugago}
I.~L.~Buchbinder, S.~J.~J.~Gates, H.~S.~J.~Goh, W.~D.~.~Linch, M.~A.~Luty, S.~P.~Ng and J.~Phillips,
Phys.\ Rev.\ D {\bf 70} (2004) 025008
[arXiv:hep-th/0305169].

\bibitem{rascst}
R.~Rattazzi, C.~A.~Scrucca and A.~Strumia,
Nucl.\ Phys.\ B {\bf 674} (2003) 171
[arXiv:hep-th/0305184].

\bibitem{grrasc}
T.~Gregoire, R.~Rattazzi, C.~A.~Scrucca, A.~Strumia and E.~Trincherini,
arXiv:hep-th/0411216.
C.~A.~Scrucca,
arXiv:hep-th/0412237.


\bibitem{alle}
Z.~Lalak and R.~Matyszkiewicz,
Phys.\ Lett.\ B {\bf 583} (2004) 364
[arXiv:hep-th/0310269].
N.~Maru and N.~Okada,
Phys.\ Rev.\ D {\bf 70} (2004) 025002
[arXiv:hep-th/0312148].
H.~Abe, K.~Choi and I.~W.~Kim,
JHEP {\bf 0409} (2004) 001
[arXiv:hep-th/0405100].
P.~Brax and N.~Chatillon,
Phys.\ Lett.\ B {\bf 598} (2004) 99
[arXiv:hep-th/0407025].
F.~P.~Correia, M.~G.~Schmidt and Z.~Tavartkiladze,
arXiv:hep-th/0410281.
G.~von Gersdorff, L.~Pilo, M.~Quiros, A.~Riotto and V.~Sanz,
arXiv:hep-th/0411133.
H.~Abe and K.~Choi,
JHEP {\bf 0412} (2004) 069
[arXiv:hep-th/0412174].
F.~P.~Correia, M.~G.~Schmidt and Z.~Tavartkiladze,
arXiv:hep-th/0501209.

\bibitem{zu}
M.~Zucker,
Phys.\ Rev.\ D {\bf 64} (2001) 024024
[arXiv:hep-th/0009083].
M.~Zucker,
Fortsch.\ Phys.\  {\bf 51}, 899 (2003).

\bibitem{fukuoh}
T.~Fujita, T.~Kugo and K.~Ohashi,
Prog.\ Theor.\ Phys.\  {\bf 106} (2001) 671
[arXiv:hep-th/0106051].
T.~Kugo and K.~Ohashi,
Prog.\ Theor.\ Phys.\  {\bf 108} (2002) 203
[arXiv:hep-th/0203276].

\bibitem{howi}
P.~Horava and E.~Witten,
Nucl.\ Phys.\ B {\bf 475} (1996) 94
[arXiv:hep-th/9603142].

\bibitem{aa_master}
 A. Falkowski, {\it M.Sc. Thesis} (2000) \\ 
 http://info.fuw.edu.pl/\~{}afalkows/Pub/publikacje.html

\bibitem{chni}
A.~H.~Chamseddine and H.~Nicolai,
Phys.\ Lett.\ B {\bf 96} (1980) 89.
M.~Gunaydin, G.~Sierra and P.~K.~Townsend,
Nucl.\ Phys.\ B {\bf 242} (1984) 244.

\bibitem{mipe}
E.~A.~Mirabelli and M.~E.~Peskin,
Phys.\ Rev.\ D {\bf 58} (1998) 065002
[arXiv:hep-th/9712214].

\bibitem{digeko}
G.~A.~Diamandis, B.~C.~Georgalas, P.~Kouroumalou and A.~B.~Lahanas,
Phys.\ Lett.\ B {\bf 602} (2004) 112
[arXiv:hep-th/0402228].

\bibitem{ceag}
M.~Gunaydin, G.~Sierra and P.~K.~Townsend,
Nucl.\ Phys.\ B {\bf 253} (1985) 573.
M.~Gunaydin and M.~Zagermann,
Nucl.\ Phys.\ B {\bf 572} (2000) 131
[arXiv:hep-th/9912027].
A.~Ceresole and G.~Dall'Agata,
Nucl.\ Phys.\ B {\bf 585} (2000) 143
[arXiv:hep-th/0004111].
E.~Bergshoeff, S.~Cucu, T.~de Wit, J.~Gheerardyn, S.~Vandoren and A.~Van Proeyen,
Class.\ Quant.\ Grav.\  {\bf 21} (2004) 3015
[arXiv:hep-th/0403045].

\bibitem{ghpo}
T.~Gherghetta and A.~Pomarol,
Nucl.\ Phys.\ B {\bf 586} (2000) 141
[arXiv:hep-ph/0003129].
A.~Falkowski, Z.~Lalak and S.~Pokorski,
Phys.\ Lett.\ B {\bf 491}, 172 (2000)
[arXiv:hep-th/0004093].
A.~Falkowski, Z.~Lalak and S.~Pokorski,
Phys.\ Lett.\ B {\bf 509}, 337 (2001)
[arXiv:hep-th/0009167].

\bibitem{rasu}
L.~Randall and R.~Sundrum,
Phys.\ Rev.\ Lett.\  {\bf 83} (1999) 3370
[arXiv:hep-ph/9905221].


\bibitem{babe}  
J.~Bagger and D.~V.~Belyaev,
Phys.\ Rev.\ D {\bf 67}, 025004 (2003)
[arXiv:hep-th/0206024].

\bibitem{babe2}
J.~A.~Bagger and D.~V.~Belyaev,
arXiv:hep-th/0406126.

\bibitem{crjusc}
E.~Cremmer, B.~Julia, J.~Scherk, S.~Ferrara, L.~Girardello and P.~van Nieuwenhuizen,
Nucl.\ Phys.\ B {\bf 147} (1979) 105.

\bibitem{popo}
E.~Ponton and E.~Poppitz,
JHEP {\bf 0106} (2001) 019
[arXiv:hep-ph/0105021].

\bibitem{dvgapo}
G.~R.~Dvali, G.~Gabadadze and M.~Porrati,
Phys.\ Lett.\ B {\bf 485} (2000) 208
[arXiv:hep-th/0005016]. 

\bibitem{raka}
A.~Karch and L.~Randall,
JHEP {\bf 0105} (2001) 008
[arXiv:hep-th/0011156].
\end{thebibliography}
\end{document}